\documentstyle[aps,prl,epsf]{revtex}
\newcommand{\beq}{\begin{equation}}
\newcommand{\eeq}{\end{equation}}
\newcommand{\beqa}{\begin{eqnarray}}
\newcommand{\eeqa}{\end{eqnarray}}
\newcommand{\omn}{\omega_n}
\newcommand{\ko}{K_{\rho 1}}

\newcommand{\kt}{K_{\rho 2}}

\newcommand{\koo}{K_{\rho 0}}
\newcommand{\ksoo}{K_{\sigma 0}}
\newcommand{\up}{\uparrow}
\newcommand{\down}{\downarrow}
\newcommand{\bi}{\bibitem}
\newcommand{\la}{\label}

\newcommand{\vo}{v_{\rho 1}}
\newcommand{\vt}{v_{\rho 2}}

\begin{document}
\draft
\twocolumn[\hsize\textwidth\columnwidth\hsize\csname@twocolumnfalse%
\endcsname
\title{Quasi-Andreev reflection in inhomogeneous Luttinger liquids}
\author{Dmitrii L. Maslov}
\address{
Department of Physics, University of Florida,
P.\ O.\ Box 118440, Gainesville, Florida 32611}
\author{Paul M. Goldbart}
\address{
Department of Physics and Materials Research Laboratory,
University of Illinois at Urbana-Champaign,
Urbana, Illinois 61801}
\maketitle
\begin{abstract}
Reflection of charge excitations at the step
in the interaction strength in a Luttinger liquid
can be of the Andreev type, even the interactions
are purely repulsive. The region with
stronger repulsion plays the role of a normal
metal in a normal-metal/superconductor junction,
whereas the region with weaker repulsion plays the
role of a superconductor. It is shown that
this quasi-Andreev reflection leads to a number
of proximity-like effects, including the local enhancement
(suppression) of  superconducting fluctuations
on the quasi-normal (quasi-superconducting) side
of the step, significant modification of the local density
of states, as well as others. The observable 
consequences of these proximity effects
are analyzed for the case of 
single- and two-particle
tunneling from a normal-metal
or superconducting tip into an inhomogeneous
Luttinger-liquid wire.  
\end{abstract}
]
Recent developments in microfabrication technologies
have led to a renaissance
in the physics of (quasi) one-dimensional (1D)
strongly correlated electronic systems. 
Much of theoretical
work has been devoted to various scattering
processes in these systems, such as single- \cite{single} and multiple- \cite{multiple} impurity scattering, and Umklapp scattering
\cite{umklapp}. These processes result
in some unique features in observable
quantities, e.g., in specific temperature
and/or voltage dependences of the conductance,
which have been instrumental in the experimental
search for strongly correlated effects in 1D systems \cite{exp}.

The focus of this paper is on 
another scattering process, which has
only recently attracted attention
\cite{oreg,safi,raikh,chklovskii,sandler},
namely, scattering caused by 
{\it inhomogeneities} in the electron-electron interaction strength.
Such inhomogeneities should be readily realizable,
and sometimes even unavoidable,
in 1D systems. For example, one can
change the electron density, and therefore
the effective interaction strength,  by applying
a potential to a top gate that covers
a part of a quantum wire (see Fig.~\ref{fig:fig1}a).
Squeezing a wire inhomogeneously by using
additional side gates
(Fig.~\ref{fig:fig1}b)
has the same effect:
the effective 1D interaction constant $U_0$
is proportional to $1/w(x)$, where $w(x)$ is the
local width of the wire \cite{fn_2D}. 
Such inhomogeneities can (and should)
exist in their random versions: e.g.,
disorder can cause the random modulation of the
electron density \cite{raikh}, and surface
roughness can lead to random modulation of the
width. Finally, 
such scattering should occur at an interface
between regions having different filling
factors in a quantum Hall system \cite{chklovskii,sandler}.
In what follows, we shall concentrate
on the {\it adiabatic} case, in which the scale of the
inhomogeneity $W$ is much larger than
the Fermi wavelength $2\pi/k_F$. In this case, single-particle
backscattering from the
inhomogeneity may be neglected \cite{fn_chem}.~
\begin{figure}[htb]
\setlength{\unitlength}{1.0in}
\begin{picture}(2.0,2.5)(0.1,0) 
\put(0.15,0.1){\epsfxsize=3.0in\epsfysize=2.2in\epsfbox{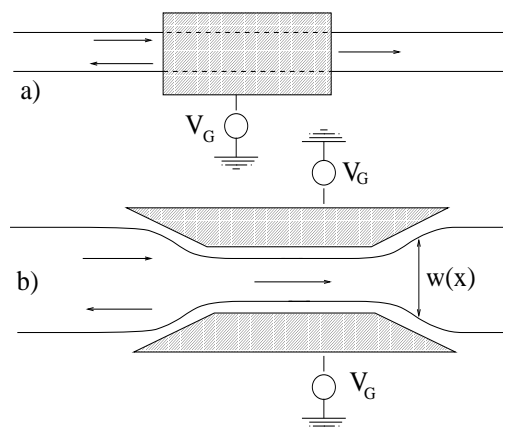}}
\end{picture}
  \caption{Some ways of producing inhomogeneities
in the interactions strength:
(a) Creating a depleted (enriched) region by applying
a voltage to a top gate (shaded);
(b) Squeezing the wire by additional side gates. As soon as
$[k_F w(x)/\pi]=1$ for any $x$, there is only one
propagating mode in the wire.}
\la{fig:fig1}
\end{figure}
Another advantage of the adiabatic approximation
is that a 1D system with variable interaction strength 
can be described in terms
of the inhomogeneous Luttinger-liquid model \cite{safi,ms,ponomarenko},
in which the velocity of collective
charge excitations $v_{\rho}$ and the dimensionless parameter
$K_{\rho}$  vary in space.
($K_{\rho}$ characterizes the strength and sign of interactions:
$K_{\rho}<1$ for repulsion; $K_{\rho}>1$ for attraction;
$K_{\rho}=1$ in the absence of the interaction.)
An interface between two regions having different
interaction strengths thus corresponds to
kinks in $v_{\rho}$ and $K_{\rho}$.
(We assume that the $SU(2)$ symmetry
is preserved, and hence that everywhere the velocity of spin excitations
is equal to the bare Fermi velocity $v_F$ and $K_{\sigma}=1$.)

As has recently been noticed  by a number of authors
\cite{oreg,safi,chklovskii,sandler},
scattering 
at a kink in the interaction strength%
\begin{figure}[htb]
\setlength{\unitlength}{1.0in}
\begin{picture}(2.0,1.8)(0.1,0) 
\put(0.15,0.1){\epsfxsize=3.0in\epsfysize=2.0in\epsfbox{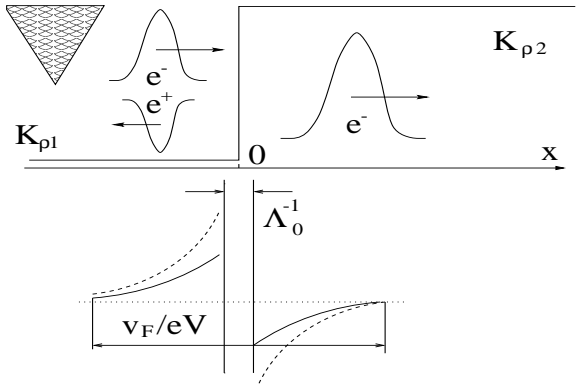}}
\end{picture}
  \caption{
(a) Quasi-Andreev reflection of an electron-like excitation at the
step in $K_{\rho}$. Shaded: tunneling tip. 
(b) Schematic view of single-particle (solid) and two-particle (dashed)
tunneling currents as a function of the position of the tip,
when the tip is in the normal and superconducting states, respectively.
The currents are normalized to their values at $x\pm\infty$ (dotted).
The region of length
$\sim \Lambda_0^{-1}\sim 1/k_F$ in the vicinity of the step,
where the long-wavelength approximation breaks
down, is excluded from the Figure.}
\la{fig:fig2}
\end{figure}
can be of the Andreev type. An important
(and readily tractable)
case occurs when the wavelength of the excitation
 $L_{\epsilon}\equiv v_F/\epsilon$
(where $\epsilon$ is a typical excitation energy and $\hbar=1$)
is much larger
that the width of the kink $W$ (but $W\gg 2\pi/k_F$). 
In this case the density reflection
coefficient $R=(\rho_{\rm out}-\rho_0)/(\rho_{\rm in}-\rho_0)$,
where $\rho_{\rm out}$ ($\rho_{\rm in}$) is
the charge density in the outgoing (incoming)
wave and $\rho_{0}$ is the background
density, is given simply by a Fresnel-type formula \cite{safi,chklovskii}
\beq
R=\frac{K_{\rho 1}-K_{\rho 2}}{K_{\rho 1}+K_{\rho 2}}.
\la{eq:refl}
\eeq
Evidently, $R<0$ for $K_{\rho 1}<K_{\rho 2}$, which means that
an incoming electron-like excitation,
corresponding to an excess in
the density (i.e., $\rho_{\rm in}>\rho_{0}$),
is reflected as a hole-like one,
corresponding to a deficit in the density
(i.e., $\rho_{out}<\rho_{0}$), and vice versa (see Fig.~\ref{fig:fig2}a).
Most importantly, this kind of Andreev
reflection, which we will refer to 
as \lq\lq quasi-Andreev reflection \rq\rq\,
occurs even in the presence of purely
{\it repulsive} interactions.
Making an analogy with a superconductor/normal
metal (SN) junction, one can say that the region
with weaker repulsion
(i.e., $x>0$) plays the role of the \lq\lq superconductor\rq\rq\
 and the region with stronger repulsion (i.e., $x<0$)  plays
the role of the \lq\lq normal metal\rq\rq, so that, loosely
speaking, \lq\lq weaker repulsion\rq\rq\/
is equivalent to \lq\lq stronger attraction\rq\rq.
To make this analogy more explicit,
we will call the regions $x>0$ and $x<0$
the \lq\lq quasi-superconducting\rq\rq\ (QS) and \lq\lq quasi-normal\rq\rq\
(QN) parts, respectively.

The goal of this paper
is to explore the features
of the electron state formed
in the vicinity of a step in the interaction strength
in a Luttinger liquid. We shall
show that quasi-Andreev reflection results
in a peculiar {\it proximity effect\/}:
 the superconducting
fluctuations are enhanced (suppressed) on the QN
(QS) side of a step. The single-particle density
 of states (DOS) will also be shown to be modified
in a similar way. We shall then
study how both of these effects
could manifest themselves in
single- and two-particle tunneling
into a Luttinger liquid.
For the sake of simplicity we shall concentrate
on the case of a single kink of zero width
(a step). Although kinks usually come in pairs, 
our calculations are applicable if the separation
of the kinks is larger than $L_{\varepsilon}$. The overall structure
of the DOS in the case of a well in $K_{\rho}$
was studied in Ref.~\cite{nazarov}.

An inhomogeneous
Luttinger liquid is described
by the Hamiltonian 
$$
\!\!\!\!H=\!\!\sum_{\mu=\rho,\sigma}\int dx \frac{v_{\mu}(x)}{2}
\Big[\frac{(\partial_{x}\phi_{\mu})^2}{K_{\mu}(x)}
+K_{\mu}(x)
(\partial_x\theta_{\mu})^2\Big],
$$
where $\phi_{\mu}$ and $\theta_{\mu}$ are canonically
conjugate boson fields, i.e., 
$[\phi_{\mu}(x),\partial_x\theta_{\mu'}(x')=i\delta_{\mu\mu'}\delta(x-x')$. 
Unless mentioned specifically, we shall be considering
the case of repulsive interactions, when there is
no gap in the spin sector. 

For a conventional SN interface,
Andreev reflection provides a microscopic mechanism
for the proximity effect, i.e., for
the formation of a superconducting condensate
in a region of the N-side adjacent to the
interface, and the suppression of the condensate
on the S-side. The proximity effect
is usually described by the profile
of the (singlet) condensate amplitude
$\langle \psi_{\uparrow}^{}(x)\psi_{\downarrow}^{}(x)\rangle$, 
which varies from some non-zero value in the bulk of S to
zero in the bulk of N. In our case, both the singlet
and triplet condensate amplitudes
are equal to zero.
However, one can expect quasi-Andreev reflection to modify
the  correlation
functions of singlet and triplet superconducting fluctuations, defined
as $F_s^{}(x,x')\equiv\langle S^{}(x)S^{\dagger}(x')\rangle$ and
$F_t^{}(x,x')\equiv\langle T^{}(x)T^{\dagger}(x')\rangle$, where
$S(x)=\sum_{\sigma=\up,\down}R_{\sigma}L_{-\sigma}$
and 
$T(x)=\sum_{\sigma=\up,\down}R_{\sigma}L_{\sigma}$,
and where $R_{\sigma}(L_{\sigma})$ are the right- (left-)moving
components of fermion fields.
In a bosonized form, the equal-time correlators
are given by
\begin{mathletters}
\beqa
F_s(x,x')&&=\frac{1}{(2\pi\alpha)^2}
e^{2\pi \left\{\Delta\Phi_{\sigma}(x,x',0)+\Delta\Theta_{\rho}(x,x',0)\right\}},
\la{eq:fs}\\
F_t(x,x')&&=
\frac{1}{(2\pi\alpha)^2}
e^{2\pi\left\{\Delta\Theta_{\sigma}(x,x',0)+\Delta\Theta_{\rho}(x,x',0)\right\}},
\la{eq:ft}
\eeqa
\end{mathletters}
where 
$$\Delta\Phi_{\mu}(x,x',\tau)\!\equiv\!\Phi_{\mu}(x,x',\tau)\!-\!\frac{
\Phi_{\mu}(x,x,0)\!+\!\Phi_{\mu}(x',x',0)}{2},$$
with the Matsubara propagator defined as
$\Phi_{\mu}(x,x',\tau)\equiv\langle T_{\tau}\phi_{\mu}(x,\tau)
\phi_{\mu}(x',0)\rangle$ (and similarly for the relation
between $\Delta\Theta_{\mu}$, $\Theta_{\mu}$, and $\theta_{\mu}$),
and $\alpha$ is a short-distance cutoff.
In the presence of an inhomogeneity, the temporal Fourier transform
of $\Phi_{\mu}$ satisfies the equation of motion
\beq
\Big\{ \frac{\omn^2}{v_{\mu}(x)K_{\mu}(x)}-
\partial_x\left(\frac{v_{\mu}(x)}{K_{\mu}(x)}\partial_x
\right)\!\!\Big\}\Phi_{\mu}=\delta(x-x'),
\la{eq:eqmot}
\eeq
which, in the case of a step in $v_{\rho}$ and $K_{\rho}$,
is supplemented by
the condition that $\Phi_{\mu}$ and $(v_{\mu}/K_{\mu})\partial_x\Phi_{\mu}$
are continuous
at $x=0$ \cite{ms}. The equation of motion and the boundary condition
for $\Theta_{\mu}$ are obtained by the replacement
$K_{\mu}\to 1/K_{\mu}$. In the presence
of the $SU(2)$ symmetry, the spin propagators
are given by $\Phi_{\sigma}=\Theta_{\sigma}= \exp(-|\omn(x-x')|/v_F)/2|\omn|$,
whereas $\Theta_{\rho}$ for $x,x'<0$ can be written as
\beq
\Theta_{\rho}(x,x',\omn)=\frac{
e^{-|\omn(x-x')|/v_{\rho 1}}+R
e^{-|\omn(x+x')|/v_{\rho 1}}}{2|\omn|\ko}
\la{eq:theta}
\eeq
At zero temperature, the superconducting correlation
functions take the following form
\beq
F_s(x,x')\!=F_t(x,x')\!
=f^{\{0\}}(x-x')\left[\frac{4xx'}{(x+x')^2}\right]^{\frac{R}{2\ko}}\!\!,
\la{eq:FFstep}
\eeq
where $f^{\{0\}}(x-x')\propto |x-x'|^{-(\ko^{-1}+1)}$
describes the decay of superconducting fluctuations
in a homogeneous Luttinger liquid with parameter
$\ko$ \cite{emery}.
The presence
of the inhomogeneity is manifested
through the factor in the square brackets in Eq.~(\ref{eq:FFstep}).
For the case shown in Fig.~\ref{fig:fig2}b, $R<0$ and
$F_{s/t}$ diverges as $|x|^{-|R|/2\ko}$ for fixed $x'$ and 
$x\to 0^{-}$.
For $x,x'>0$, the propagator
$\Theta_{\rho}$ is obtained from Eq.~(\ref{eq:theta})
by replacing $\ko\to\kt$, $\vo\to\vt$. 
Consequently,
$F_{s/t}$ vanishes as $x^{|R|/2\kt}$ for fixed $x'$ and $x\to 0^{+}$.
We thus see that
the superconducting fluctuations are enhanced (suppressed) on
the QN (QS) side. This is not
the only consequence of the inhomogeneity however.
As one can show, the charge- and spin-density-wave
(CDW and SDW) fluctuations
also get modified in a manner opposite to that
of the superconducting fluctuations, i.e.,
the $2k_F$ components of CDW and SDW correlation functions
vanish (diverge) as $|x|^{|R|\ko/2}$ ($|x|^{-|R|\kt/2}$)
for $x$ approaching the step from the QN
(QS) side. Last, but not least, the local
single-particle 
DOS at fixed energy $\epsilon$
diverges as $|x|^{-\kappa_{1}}$ for $x\to 0^{-}$,
where $\kappa_{1}=\frac{1}{4}|R|(\ko^{-1}-\ko^{})$
and vanishes as $x^{\kappa_{2}}$ for $x\to 0^{+}$, where
$\kappa_{2}=\frac{1}{4}|R|(\kt^{-1}-\kt^{})$;
the bulk behavior is restored at distances
$|x|\gg L_{\epsilon}$.

The modification of the DOS and various two-particle
correlation functions should lead
to some observable consequences.
In what follows, we consider
a particular example, namely, we
study how  tunneling
between a Luttinger liquid and a normal-metal
or superconducting tip
is modified in the presence of an inhomogeneity.
Tunneling from a normal metal tip
measures the single-particle DOS of a Luttinger liquid.
Tunneling between a normal-metal
(including Luttinger-liquid) conductor and a superconductor
at biases smaller than the
superconducting energy gap measures
the pair susceptibility
of a normal metal \cite{scalapino},
which is related to the correlation
function of superconducting fluctuations.
The CDW- and SDW-susceptibilities
do not enter the tunneling current,
as they correspond to electrically neutral
excitations.

Suppose that a narrow tunneling tip is scanned
along an inhomogeneous Luttinger liquid wire
(see Fig.\ref{fig:fig2}a). We  model
the tip by a 1D Luttinger liquid, whose parameters
$\koo$ and $\ksoo$ are chosen to be either
$\koo=\ksoo=1$ (when the tip is in the normal, Fermi-liquid
state) or $\koo=\infty$, $\ksoo=0$ (when
the tip is in the superconducting state). 
The dynamics of the fields in the
wire at the position of the tip $x$ is described by 
a local action obtained from the full action
by integrating out the bulk degrees of freedom
\beq
S=\frac{1}{2\beta}\sum_{\omn}\sum_{\mu=\rho,\sigma}\frac{1}{\Phi_{\mu}(x,x,\omn)}
|\phi_{\mu}(x,\omn)|^2.
\la{eq:s0phi}
\eeq
Equivalently, $S$ can be written in a dual form
by making the replacement $\phi_{\mu}\to\theta_{\mu}$
and $\Phi_{\mu}\to\Theta_{\mu}$.
The tip is described by similar equations
involving corresponding (homogeneous) propagators of boson fields.
The single-particle
tunneling action is given by
\beqa
S_1=\frac{\gamma v_F}{\alpha}\int d\tau\,&&
 \cos\left[\sqrt{\pi}\vartheta_{\rho}+{\cal A}(\tau)\right]
\cos(\sqrt{\pi}\vartheta_{\sigma})\\\nonumber
&&\times\cos(\sqrt{\pi}\varphi_{\rho})
\cos(\sqrt{\pi}\varphi_{\sigma}),
\label{eq:spt}
\eeqa
where $\gamma$ is a dimensionless tunneling amplitude, 
$\vartheta_{\mu}\equiv (\theta_{\mu}-\theta_{\mu 0})/\sqrt{2}$,
$\varphi_{\mu}\equiv (\phi_{\mu}-\phi_{\mu 0})/\sqrt{2}$,
subindex \lq\lq $0$\rq\rq\ denotes the boson fields
in the tip, and where ${\cal A}(\tau)=e\int^{\tau} d\tau'V(\tau')$,
with $V(\tau)$ being the bias between the tip and the wire.
As is well known, single-particle tunneling action $S_1$ generates
two-particle tunneling terms under a renormalization group (RG) procedure
\cite{yakovenko,nersesyan,khveshenko,tsvelik}. For electrons
with spin, the two-particle tunneling action was written down
by Khveshenko and Rice \cite{khveshenko,fn_fusion}
\beqa
S_2=\frac{v_F}{\alpha}\int d\tau&&
\{\Gamma_{s}\cos(2\sqrt{\pi}\vartheta_{\rho}+
2{\cal A})\cos(2\sqrt{\pi}\varphi_{\sigma})\\\nonumber
&&+\Gamma_{t}\cos(2\sqrt{\pi}\vartheta_{\rho}+2{\cal A})\cos(2\sqrt{\pi}\vartheta_{\sigma})
\\\nonumber
&&+\Gamma_{cdw}\cos(2\sqrt{\pi}\varphi_{\rho})\cos(2\sqrt{\pi}\varphi_{\sigma})
\\\nonumber
&&+\Gamma_{sdw}\cos(2\sqrt{\pi}\varphi_{\rho})\cos(2\sqrt{\pi}\vartheta_{\sigma})\}.
\label{eq:tpt}
\eeqa
The first (last) two terms on the right hand side of
 Eq.~(8) correspond
to particle-particle (particle-hole) tunneling. 
The first term
describes the tunneling of a (virtual) singlet Cooper pair 
from the tip to the wire (and vice versa).
Similarly, the second term describes the tunneling of a
(virtual) triplet Cooper pair. Finally, the third and fourth term
correspond to tunneling processes of electron-hole pairs
of CDW- and SDW-type, respectively.

The RG equations for the single- and two-particle
tunneling amplitudes can be derived following the conventional
procedure. When the tip probes, e.g., the
QN part of the wire, the flow equation
for $\gamma$ takes the form
\beq
\frac{d\ln \gamma}{d\ln \Lambda}=\frac{1}{2}\left(\kappa+\kappa_1
e^{-2\Lambda |x|}\right),
\la{eq:gammaflow}
\eeq 
where $\Lambda\equiv 1/\alpha$ and
\beq
\kappa=\frac{\ko+\ko^{-1}+\koo+\koo^{-1}+\ksoo+
\ksoo^{-1}}{4}-\frac{3}{2}.
\label{eq:kappa}
\eeq
The solution of Eq.~(\ref{eq:gammaflow}) is
\beq
\gamma(\Lambda)=\gamma_0\left(\Lambda/\Lambda_0
\right)^{\kappa/2}
e^{\frac{1}{2}\kappa_1\left\{E_1(2\Lambda_0|x|)-
E_1(2\Lambda |x|)\right\}
},
\la{eq:gammasol}
\eeq
where $\Lambda_0\sim k_F$ and $E_1(y)=\int^{\infty}_{1}dt\, e^{-yt}/t$. If the
tip is in the normal state ($\koo=\ksoo=1$) then
at distances $x$ from the step satisfying $\Lambda_0^{-1}\ll |x|\ll 1/\Lambda$
Eq.~(\ref{eq:gammasol}) reduces to
\beq
\gamma(\Lambda)\propto \Lambda^{\left(\ko-1\right)^2/8\ko} \left(\Lambda|x|\right)^{-|\kappa_1|/2}.
\la{eq:gasympt}
\eeq
The (dimensionless) tunneling
conductance $G_1\sim\gamma^2(\Lambda=eV/v_F)$ is thus enhanced in the vicinity
of the step and reverts to its
bulk value for $|x|\gg v_F/eV$. Accordingly, $G_1$ is suppressed if the tip
probes the QS region.
The voltage dependence of the conductnce
can be read off from the energy dependence
of the DOS: the first factor in Eq.~(\ref{eq:gammasol})
corresponds to the DOS of a homogeneous Luttinger liquid, while
the second one arises due to the step.
 If the tip is in the superconducting
state ($\koo=\infty$, $\ksoo=0$), $\gamma$ vanishes for $\Lambda\to 0$,
reflecting the absence of a single-particle tunneling
current between a normal metal and a superconductor
at voltages below the superconducting gap.

Next we consider the renormalization of the two-particle
tunneling amplitudes. The charge transfer between
the tip and the wire is determined only by the particle-particle
tunneling processes, and we thus need to consider only the flow
equations for $\Gamma_{s}$ and $\Gamma_{t}$, which take the form
\beq
\frac{d\Gamma_{i}}{d\ln\Lambda}=\frac{1}{2}
\left(\kappa_{i}+\frac{R}{\ko}e^{-2\Lambda|x|)}
\right)\Gamma_{i}+\lambda_{i}\gamma^2(\Lambda),
\la{eq:tflow}
\eeq
where $i=s,t$ and
\begin{mathletters}
\beqa
&\kappa_{s}=\ko^{-1}+\koo^{-1}+\ksoo-1;\\\nonumber
&\lambda_{s}=\ko-\ko^{-1}+\koo-\koo^{-1}+\ksoo^{-1}-\ksoo\\\la{eq:kapss}
&-c\left(\ko^{-1}+\ko\right); \;c\sim 1.\la{eq:lamss}
\eeqa
\end{mathletters}%
The parameters $\kappa_{t}$ and $\lambda_{t}$ are obtained
from $\kappa_{s}$ and $\lambda_{s}$ by replacing
$\ksoo^{}\to\ksoo^{-1}$. The first term in Eq.~(\ref{eq:tflow})
describes the self-generation of two-particle tunneling, the second
one represents the generation of two-particle tunneling
by single-particle processes. Accordingly, the
initial condition for $\Gamma_i$ is $\Gamma_{i}(\Lambda_0)=0$
 (if the tip is normal)
and $\Gamma(\Lambda_0)=t_{0i}\neq 0$ (if
the tip is superconducting). Analysis of the former
case shows that two-particle tunneling does not
 significantly modify the tunneling conductance
for repulsive interactions: two-particle
tunneling is irrelevant and subdominant to
single-particle tunneling.

The situation changes dramatically
when the tip is superconducting. In this
case, single-particle tunneling is forbidden
at voltages smaller then the superconducting
gap, and the only current flowing from
the tip to the wire is the two-particle
one. Solving Eq.~(\ref{eq:tflow})
with $\gamma(\Lambda)=0$, $\koo=\infty$,  and $\ksoo=0$,
we get for $\Gamma_{s}$
\beq
\Gamma_{s}(\Lambda)=\Gamma_0\left(\Lambda/\Lambda_0
\right)^{q_s/2}
e^{\frac{R}{2\ko}\left\{E_1(2\Lambda_0|x|)-
E_1(2\Lambda |x|)\right\}
},
\la{eq:tsol}
\eeq
where $q_s\equiv \ko^{-1}-1$, 
whereas $\Gamma_{t}=0$ (there is no triplet tunneling
into a singlet superconductor). Close to
the step, the (dimensionless) tunneling conductance $G_2\sim \Gamma^2_{s}(\Lambda=eV/v_F)$ is
\beq
G_2\propto t_{0s}^2V^{q_{s}}\left(V|x|\right)^{-|R|/\ko}.
\la{eq:conds}
\eeq
In a homogeneous wire ($R=0$), $G\propto V^{q_{s}}$. Two-particle
tunneling is irrelevant for the case of repulsion ($\ko<1$)
and relevant for attraction ($\ko>1$), the non-interacting
case ($\ko=1$) being marginal. The step enhances (suppresses) the
two-particle tunneling for a tip to the left (right) from
the step. The enhancement can be so strong that
two-particle tunneling becomes relevant, even
for repulsion: the criterion of relevancy
changes to $\ko>1-|R|$, which can be satisfied 
for $\ko<1$. However, if the reflection coefficient
is given by Eq.~(\ref{eq:refl}), this criterion
can be satisfied only if the interaction in
the second part of the wire is attractive
(i.e., $\kt>2-\ko$).

We thank O.\ A.\ Starykh for
numerous illuminating discussions. D.\ L.\ M.\ is grateful
to V.\ B.\ Geshkenbein for useful comments.
This work was supported by the NSF DMR-9703388 
and the University of Florida (D.\ L.\ M.\ )
and by the DOE DEFG02-96ER45439 (P.\ M.\ G.\ ).


\begin{references}
\bi{single}C.\ L.\ Kane and M.\ P.\ A.\ Fisher,
 \prl {\bf 68}, 1220 (1992); \prb
{\bf 46}, 15233 (1992).
\bi{multiple}W.\ Apel and T.\ M.\ Rice, \prb {\bf 26}, 
7063 (1982).
\bi{umklapp} T. Giamarchi, \prb {\bf 44}, 2905 (1991).
\bi{exp} F.\ P.\ Milliken et al.,
 Sol.\ State 
Commun.\ {\bf 97}, 309 (1996); S.\ Tarucha et al.,
{\sl ibid.} {\bf 94}, 413 (1995); A.\ Yacoby et al.,
\prl {\bf 77}, 4612 (1996); A.\ Chang et al.,
{\sl ibid.} {\bf 77}, 2538 (1997).
\bi{fn_2D}A limiting case $w(x)\to\infty$
and, consequently, $U_0\to 0$
corresponds to an adiabatic contact with a 2D reservoir,
considered in Refs.~\cite{safi,ms,ponomarenko}.
\bi{oreg} Y.\ Oreg and A.\ M.\ Finkel'stein,
\prl {\bf 74}, 3668 (1995).
\bibitem{safi} I.\ Safi and H.\ J.\ Schulz,
\prb {\bf 52}, R17040 (1995);
in {\it Quantum Transport in Semiconductor Heterostructures}, 
ed. B. Kramer (Kluwer, Dordrecht, 1995); in: {\it Correlated Fermions
and Transport in Mesoscopic Systems}, ed. T.\ Martin et al.
(Editions Frontieres, Gif-sur-Yvette), p. 37 (1996).
\bi{raikh} A.\ Gramada, M.\ Raikh, \prb {\bf 55}, 7673 (1997).
\bi{chklovskii}D.\ B.\ Chklovskii and B.\ I.\ Halperin,
preprint (cond-mat/9612127).
\bi{sandler}N.\ P.\ Sandler, C.\ de C.\ Chamon and E.\ Fradkin,
preprint (cond-mat/9704012).
\bi{fn_chem}The interaction-induced
variation of the chemical
potential can be added to the
single-particle potential,
and thus neglected altogether.
\bibitem{ms} D.\ Maslov, M.\ Stone,
\prb {\bf 52}, R5539 (1995).
\bibitem{ponomarenko}
V.\ V.\ Ponomarenko, \prb {\bf 52}, R8666 (1995).
\bi{nazarov} Yu. V. Nazarov, A.\ A.\ Odintsov, and D.\ V.\ Averin,
Bull.\ Am.\ Phys.\ Soc. {\bf 41}, 376 (1996).
\bi{emery}V.\ J.\ Emery, in {\sl Highly Conducting One-Dimensional
Solids}, edited by J.\ T.\ Devreese (Plenum, New York, 1979), p. 327.
\bi{scalapino} D.\ J.\ Scalapino, \prl {\bf 24}, 1052 (1972).
\bi{khveshenko} D.\ Khveshenko, T.\ M.\ Rice, \prb {\bf 50}, 252 (1994).
\bi{yakovenko} S.\ A.\ Brazovskii and V.\ M.\ Yakovenko, 
Zh.\ Eksp.\ Teor.\ Fiz.\ {\bf 89}, 1340 (1985) [ JETP {\bf 62},
1340 (1985)];
V.\ M.\ Yakovenko, Pis'ma Zh.\ Eksp.\ Teor.\ Fiz. {\bf 56}, 523 (1992)
[JETP Lett. {\bf 56}, 5101 (1992)].
\bi{nersesyan} A.\ A.\ Nersesyan, A.\ Luther, and F.\ V.\ Kusmartsev,
 Phys.\ Lett.\ A{\bf 176}, 363 (1993). 
\bi{tsvelik} A.\ M.\ Tsvelik, {\sl Quantum Field Theory in Condensed
Matter Physics} (Cambridge University Press, 1995).
\bi{fn_fusion}The two-particle tunneling
action is not affected by inhomogeneity in the wire. Indeed, the two-particle
tunneling terms occur as a result of fusions of fermion
fields \cite{tsvelik}, which initially were evaluated
at different space-time points. Fusion reduces the separation
between these points down to the scale of 
a microscopic cutoff. At such
distances, the presence of a smooth (on a microscopic
scale) inhomogeneity is irrelevant.
\end{references}
\end{document}